\documentclass[prd,amsmath,twocolumn,10pt,superscriptaddress,floatfix,nofootinbib]{revtex4-2}
\usepackage{epsfig,amssymb,amsfonts,amsmath,mathtools,bm,color,xcolor,graphicx,braket,adjustbox,esint,upgreek}
\usepackage{multirow}
\usepackage[utf8]{inputenc}
\usepackage{setspace}
\usepackage{dcolumn,kantlipsum}
\allowdisplaybreaks

\usepackage[
colorlinks=true,
linkcolor=blue,
breaklinks=true,
urlcolor=blue,
citecolor=blue]{hyperref}

\usepackage{orcidlink}
\usepackage{mathrsfs} 

\usepackage{bm,bbm}
\usepackage{slashed}

\newcommand{\be}{\begin{equation}}
\newcommand{\ee}{\end{equation}}
\newcommand{\bea}{\begin{eqnarray}}
\newcommand{\eea}{\end{eqnarray}}
\newcommand{\beas}{\begin{eqnarray*}}
\newcommand{\eeas}{\end{eqnarray*}}

\def\vec#1{\boldsymbol{#1}}
\newcommand{\nn}{\nonumber}

\newcommand{\GeV}{\,\text{GeV}}
\renewcommand{\vec}[1]{\mathbf{#1}}

\newcommand{\ustb}{\affiliation{School of Mathematics and Physics, University
of Science and Technology Beijing, Beijing 100083, China}}

\newcommand{\uestc}{\affiliation{School of Physics, University of Electronic Science and Technology of China, Chengdu 611731, China}}

\newcommand{\itp}{\affiliation{Institute of Theoretical Physics, Chinese Academy of Sciences, Beijing 100190, China}}

\newcommand{\ucas}{\affiliation{School of Physical Sciences, University of Chinese Academy of Sciences, Beijing 100049, China}}

\newcommand{\peng}{\affiliation{Peng Huanwu Collaborative Center for Research and Education, Beihang University, Beijing 100191, China}}

\graphicspath{{FiguresV2/}}

\begin{document}

\title{Determination of the $Z_c(3900)$ and the $Z_{cs}(3985)$ states from joint analysis of experimental and lattice data}

\author{Yun-Hua Chen\orcidlink{0000-0001-8366-2170}}\email{ yhchen@ustb.edu.cn}
\ustb

\author{Meng-Lin Du\orcidlink{0000-0002-7504-3107}}\email{ du.ml@uestc.edu.cn}
\uestc

\author{Feng-Kun Guo\orcidlink{0000-0002-2919-2064}}\email{ fkguo@itp.ac.cn}
\itp \ucas \peng


\begin{abstract}
We present a unified analysis of the $Z_c(3900)$ and $Z_{cs}(3985)$ states considering both experimental and lattice data. The study simultaneously includes the processes $e^+e^- \rightarrow J/\psi \pi^+\pi^-, J/\psi K^+ K^-, D^0 D^{\ast-} \pi^+, (D^{\ast 0} D_s^{-}+D^0 D_s^{\ast -}) K^+$, together with finite-volume energy levels from recent lattice QCD simulations.
Open-charm meson loops with triangle singularities, the $J/\psi\pi(J/\psi \bar{K})$-$\bar{D}D^*(\bar{D}D^*_s)$ coupled-channel interactions, and the $\pi\pi$-$K\bar K$ final-state interaction are all taken into account. We find that pole contributions associated with the $Z_c(3900)$ and $Z_{cs}(3985)$ are indispensable for describing the data.
The successful joint description of the experimental and lattice data supports the interpretation that the $Z_c(3900)$ and $Z_{cs}(3985)$ are SU(3) flavor partners within the same octet multiplet and indicates that both are resonance states. The extracted pole masses and half-widths of the $Z_c(3900)$ and the $Z_{cs}(3985)$ are $(3879.6 \pm 4.8)$~MeV and $(32.2 \pm 4.7)$~MeV, and $(3976.9 \pm 5.1)$~MeV and $(28.8 \pm 5.9)$~MeV, respectively. The ratios of the $Z_c(Z_{cs})$ couplings to the $D\bar D^*(D_s\bar{D}^\ast+D\bar{D}_s^\ast)$ and $J/\psi \pi(J/\psi K)$ channels are also determined.
A compositeness analysis indicates that, although the $D\bar D^* (D_s\bar{D}^\ast+D\bar{D}_s^\ast)$ component in the $Z_c(3900) (Z_{cs}(3985))$ state is sizable, additional components are still needed to form these exotic states.

\medskip



\end{abstract}

\maketitle
\medskip
\medskip
\medskip

\section{Introduction}

The $Z_c(3900)^\pm$ is the first hidden-charm state with at least four valence quarks that was found by two experiments, yet its nature remains intensely debated.
It was discovered by the BESIII and Belle Collaborations in 2013 through the $e^+e^-\to J/\psi\pi^+\pi^-$ process~\cite{Ablikim:2013mio,Liu:2013dau}, and was subsequently confirmed in an analysis of the CLEO-c data~\cite{Xiao:2013iha} and in semi-inclusive $b$-hadron decays at the D0 experiment~\cite{D0:2018wyb}.
A similar exotic state, $Z_c(3885)^+$, was later observed in the $(D\bar{D}^\ast)^+$ distributions from $e^+e^-\to \pi (D\bar{D}^\ast)^+$~\cite{BESIII:2013qmu,BESIII:2015pqw}. Given their nearly degenerate masses, the $Z_c(3900)$ and $Z_c(3885)$ are commonly identified as the same state and are hereafter collectively referred to as the $Z_c(3900)$.
Various models have been proposed to interpret the $Z_c(3900)$, such as the $D\bar D^*$ hadronic molecule~\cite{Wang:2013cya,Guo:2013sya, Wilbring:2013cha, He:2013nwa,Dong:2013iqa,Zhang:2013aoa, Aceti:2014uea, Albaladejo:2015lob,Albaladejo:2016jsg, Gong:2016hlt,Gong:2016jzb,He:2017lhy,Ortega:2018cnm, Du:2020vwb,Meng:2020ihj,Wang:2020dgr,Chen:2022ddj,Liu:2024nac,Wang:2023hpp}, the compact tetraquark state~\cite{Braaten:2013boa,Dias:2013xfa,Maiani:2014aja,Qiao:2013raa,Deng:2014gqa,Agaev:2017tzv,Wua:2023ntn}, or kinematic effects, including threshold cusps from coupled channels~\cite{Chen:2013wca,Swanson:2014tra,HALQCD:2016ofq} or triangle-singularity (TS) mechanisms~\cite{Wang:2013cya,Szczepaniak:2015eza,Albaladejo:2015lob,Pilloni:2016obd,vonDetten:2023uja,Nakamura:2023obk,vonDetten:2024eie,Yu:2024sqv}. For recent reviews, see Refs.~\cite{Chen:2016qju,Hosaka:2016pey,Lebed:2016hpi,Esposito:2016noz,Guo:2017jvc,Ali:2017jda,Olsen:2017bmm,Karliner:2017qhf,Yuan:2018inv,Liu:2019zoy,Brambilla:2019esw,Guo:2019twa,JPAC:2021rxu,Meng:2022ozq,Liu:2024uxn,Chen:2024eaq, Wang:2025sic, Dai:2026fkg}.
In particular, it has been realized~\cite{Guo:2020oqk, Guo:2019twa} that in order to identify the TS effects in the $Z_c(3900)$ signal, the analysis of the dependence of the signals on the $e^+e^-$ center-of-mass (c.m.) energy is crucial. 
Recently, the BESIII Collaboration released their measurements of the $Z_c(3900)$ in the $e^+e^-\to \pi (D\bar{D}^\ast)^+$ with $e^+e^-$ energies from 4.1271 to 4.3583 GeV~\cite{BESIII:2025qkn}, which offers the opportunity to establish the existence of the $Z_c(3900)$ as a resonance and extract its properties accurately. In order to reach such a conclusion, a thorough analysis with the TS effects taken into account is necessary.

Inspired by the discovery of the $Z_c(3900)$, lattice quantum chromodynamics (QCD) studies of the $Z_c(3900)$ have been performed by several groups~\cite{Prelovsek:2013xba,Prelovsek:2014swa,Chen:2014afa,Cheung:2016bym,HALQCD:2016ofq,Ikeda:2017mee,Cheung:2017tnt,Liu:2019gmh,CLQCD:2019npr,Sadl:2024dbd}. Notably, Refs.~\cite{Cheung:2017tnt,CLQCD:2019npr,Sadl:2024dbd} obtained precise finite-volume energy levels~\cite{Luscher:1986pf,Luscher:1990ux} by including the $J/\psi\pi$-$D\bar D^*$ coupled channels in their lattice QCD simulations. 
Although most of these lattice QCD calculations do not find a clear signal for the $Z_c(3900)$, it does not exclude the possibility of a pole, which needs to be determined from the $T$-matrix reconstructed from only finite-volume energy levels. In fact, the results in Refs.~\cite{Cheung:2017tnt,CLQCD:2019npr,Sadl:2024dbd} are analyzed in a unitarized amplitude approach in a finite volume, and are consistent with the existence of a $Z_c(3900)$ state~\cite{Yan:2023bwt,Sadl:2024dbd}. 
However, these analyses are incomplete in the sense that the corresponding continuum amplitudes used in fitting to BESIII data do not consider the TS kinematic effects neither the $\pi\pi$-$K\bar K$ final-state interaction (FSI). 
A combined analysis of the lattice  and experimental data, with the TS effects considered for the latter, is therefore necessary to provide a definitive conclusion regarding the existence and properties of the $Z_c(3900)$.

Moreover, in 2021, the BESIII Collaboration reported the first candidate for a charged hidden-charm tetraquark with strangeness, $Z_{cs}(3985)^-$, decaying into $D_s^- D^{\ast 0}$ and $D_s^{\ast -} D^{0}$, in the reaction $e^+e^-\to K^+(D^{*0}D_s^-+D^0D_s^{*-})$~\cite{BESIII:2020qkh}. The proximity of both the $Z_c(3900)$ and $Z_{cs}(3985)$ to their respective $D\bar{D}^\ast$ and $D\bar{D}_s^\ast/D^\ast\bar{D}_s$ thresholds, combined with their comparable widths, suggests that they be SU(3) flavor partners~\cite{Yang:2020nrt,Meng:2020ihj,Sun:2020hjw,Wang:2020htx,Xu:2020evn,Wang:2020dgr,Yan:2021tcp,Ortega:2021enc,Wu:2021ezz,Baru:2021ddn,Du:2022jjv,Wua:2023ntn}. 
However, 
the BESIII Collaboration did not find significant signal for the $Z_{cs}(3985)$ in the $J/\psi K^+$ channel of the $e^+e^-\to J/\psi K^+ K^-$ process at c.m. energies from 4.61 to 4.95 GeV~\cite{BESIII:2023wqy}.
Therefore, it is a fair question  to ask whether the two BESIII measurements are consistent with each other or not.

In this work, we solve the above two issues by performing a joint analysis of the experimental (from BESIII) and lattice data for both the $Z_c(3900)$ and $Z_{cs}(3985)$ within a unified framework, allowing a more precise determination of their properties. 


\section{Data and theoretical framework} 
\label{theor}
\begin{figure}[tb]
\centering
\includegraphics[width=\linewidth]{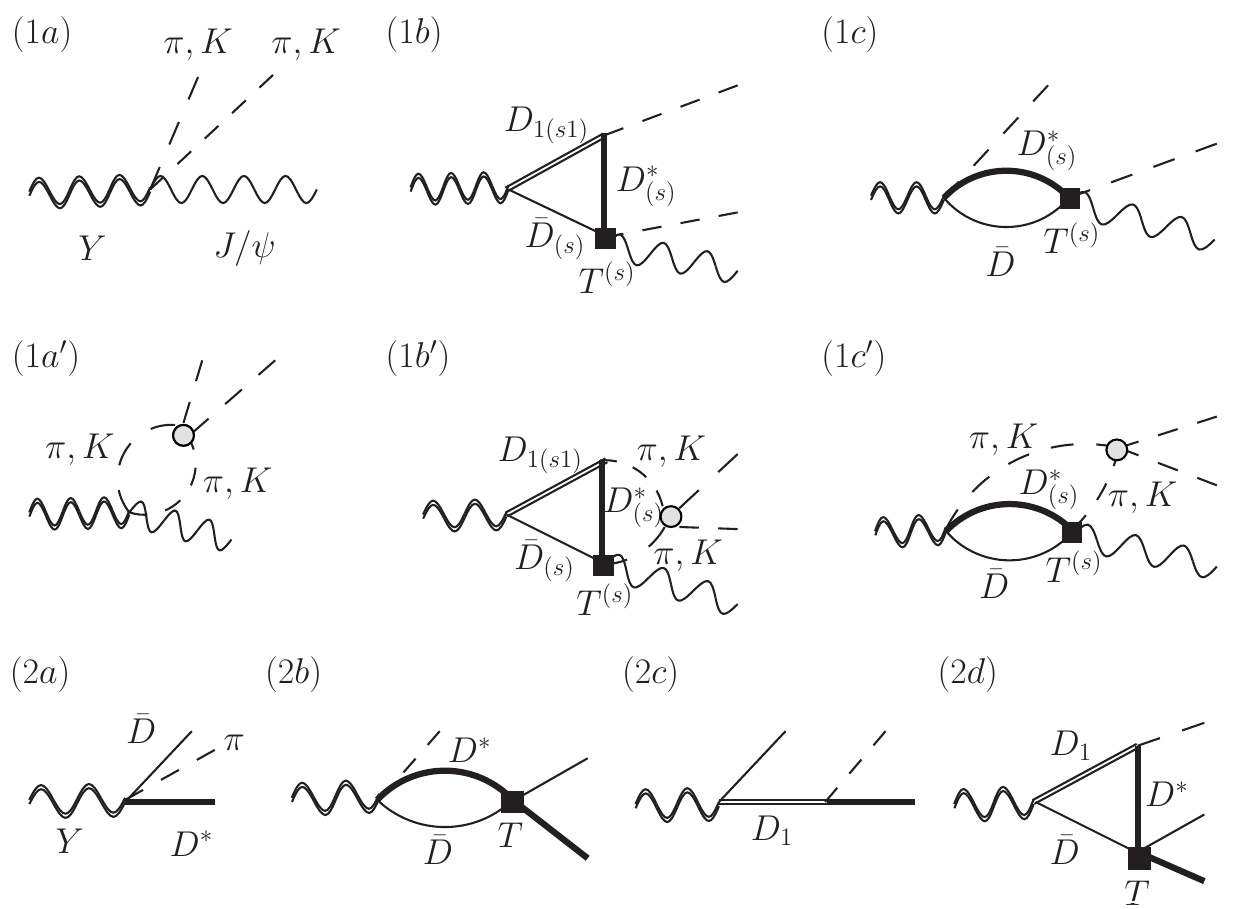}
  \caption{Diagrams for $Y \to J/\psi \Phi \Phi $ and $Y \to D^{\ast}\bar{D} \pi $. For the former: (1a) and ($\text{1a}^\prime$) represent the $Y\psi\Phi\Phi$ contact terms; (1b) and ($\text{1b}^\prime$) denote the triangle diagrams; (1c) and ($\text{1c}^\prime$) correspond to the bubble diagrams, where the primed ones denote contributions with $\pi\pi$-$K\bar K$ FSI.
  For the latter, (2a), (2b), (2c), and (2d) represent the contribution of the $Y D^\ast\bar{D}\phi$ chiral contact term, the bubble diagram, the $D_1$-exchange term, and the triangle diagram, respectively.
  The double-wavy line, dashed line, and wavy lines denote the $Y$, $\pi(K)$, and $J/\psi$ mesons, respectively. The solid, thick solid, and double-solid lines represent the $\bar D_{(s)}$, $D_{(s)}^*$, $D_{1(s1)}$ mesons, respectively.
  The gray blob represents the $\pi\pi$-$K\bar{K}$ FSI and the black square corresponds to the $J/\psi\pi(J/\psi \bar{K})$-$\bar{D}D^*(\bar{D}D^*_s)$ unitary scattering amplitude $T^{(s)}$.
  The charge conjugated charm loops are not shown explicitly but are considered in the calculation.
    }
  \label{fig.FeynmanDiagram}
\end{figure}

\begin{figure}[tb]
\centering
\includegraphics[width=\linewidth]{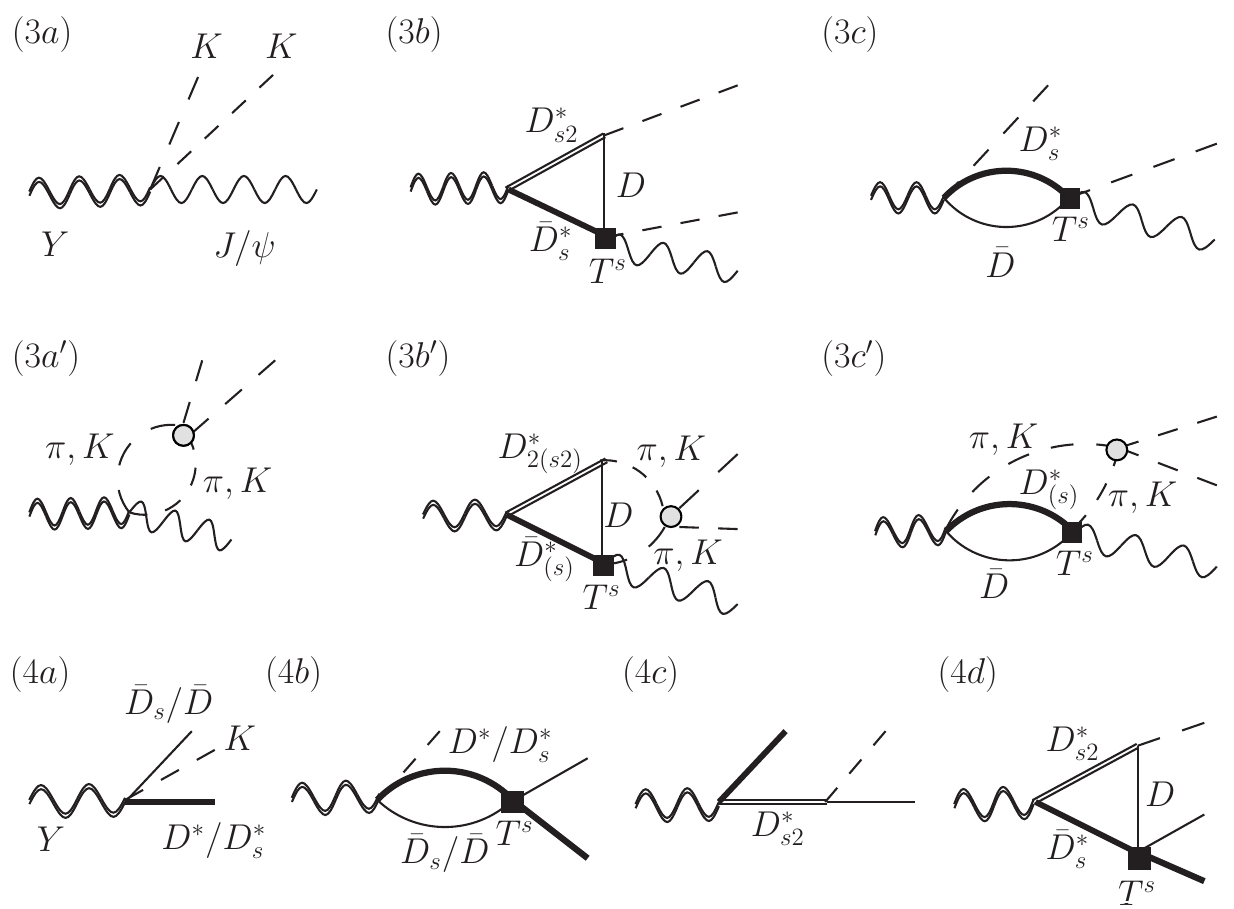}
  \caption{Diagrams for $Y \to J/\psi KK $ and $Y \to (D^{\ast}\bar{D}_s+D\bar{D}_s^\ast) K $ at $E =4.68\GeV$. The meanings of the notations are the same as in Fig.~\ref{fig.FeynmanDiagram}.
    }
  \label{fig.FeynmanDiagram4680}
\end{figure}

Our analysis includes the BESIII data sets for the $\pi^+\pi^-$ and $J/\psi\pi^\pm$ invariant mass spectra for
$e^+e^- \to J/\psi \pi^+\pi^-$~\cite{BESIII:2025qkn}, the $K^+ K^-$ invariant mass spectra for
$e^+e^- \to J/\psi K^+ K^-$~\cite{BESIII:2022joj} at 17 $e^+e^-$ c.m. energies from $E=4.1271$ to
$4.3583$~GeV, the $D^0 D^{\ast-}$ mass spectra for $e^+e^-
\rightarrow D^0 D^{\ast-} \pi^+$~\cite{BESIII:2015pqw} at $E=4.23$~GeV and $4.26$~GeV, the $K^+$ recoil-mass distribution for $e^+e^- \rightarrow (D^{\ast 0} D_s^{-}+D^0 D_s^{\ast -}) K^+$~\cite{BESIII:2020qkh}, and the $K^+K^-$ and $J/\psi K^\pm$ invariant mass spectra for
$e^+e^- \to J/\psi K^+ K^-$~\cite{BESIII:2023wqy} at $E=4.68$~GeV. We do not consider the LHCb measurements of $
B^{+} \rightarrow J / \psi \phi K^{+}
$, reporting a $Z_{cs}(4000)$, in Ref.~\cite{LHCb:2021uow}, which incolves more complicated coupled-channel dynamics~\cite{Dong:2021juy}.

We consider open-charm triangle diagrams, accounting for possible TS kinematic effects, $\pi\pi$-$K\bar K$ FSI, and $J/\psi\pi$-$D\bar D^*$ $S$-wave coupled-channel interactions (see Appendix~\ref{supp:Ts_matrix}; details for the whole formalism for $Z_c(3900)$ can be found in Ref.~\cite{Chen:2023def}, and that for the $Z_{cs}$ is similar).
The diagrams relevant for $Z_c(3900)$ and $Z_{cs}(3985)$ are shown in Figs.~\ref{fig.FeynmanDiagram} and~\ref{fig.FeynmanDiagram4680}, respectively. Here $Y$ denotes the source coupled to the virtual photon in $e^+e^-$ annihilation.
The triangle diagrams $Y \rightarrow \bar{D}D_1/\bar{D}_s D_{s1}\rightarrow \bar{D}D^\ast\pi (\bar{D}D_s^\ast K) / \bar{D}_s D^\ast K$~\cite{Cleven:2013mka,Wang:2013hga,Albaladejo:2015lob,Guo:2020oqk,Du:2022jjv} in Fig.~\ref{fig.FeynmanDiagram} and $Y \rightarrow D^\ast D_2^\ast(2460)/D_s^\ast D_{s2}^\ast(2573)\rightarrow D^\ast D\pi / D_s^\ast D K$~\cite{Cleven:2013mka,Wang:2013hga,Albaladejo:2015lob,Guo:2020oqk,Du:2022jjv} in Fig.~\ref{fig.FeynmanDiagram4680} are evaluated within nonrelativistic effective field theory. For $e^+e^-\to J/\psi\pi^+\pi^-$ and $J/\psi K^+ K^-$, we employ a model-independent dispersive approach to describe the $\pi\pi$-$K\bar K$ coupled-channel rescattering, based on unitarity, analyticity, and crossing symmetry, namely the Muskhelishvili-Omn\`es representation of the Khuri-Treiman type~\cite{Anisovich:1996tx,Garcia-Martin:2010kyn,Kubis:2015sga,Dai:2014zta,Yao:2020bxx,Chen:2023def}. The $J/\psi\pi(J/\psi \bar{K})$-$\bar{D}D^*(\bar{D}D^*_s)$ coupled-channel interactions are treated in a Lippmann-Schwinger equation approach~\cite{Albaladejo:2015lob,Yang:2020nrt}.
 
The analysis also includes the finite-volume energy levels of the $Z_c(3900)$ from lattice QCD calculations reported by the Hadron Spectrum Collaboration~\cite{Cheung:2017tnt}, by the CLQCD Collaboration~\cite{CLQCD:2019npr}, and by M.~Sadl et al.~\cite{Sadl:2024dbd}. For the lattice QCD data, we consider the energy levels up to around the $DD^\ast$ threshold. To describe them, we employ the coupled-channel L\"uscher method following the formalism in Refs.~\cite{Doring:2011vk,Doring:2012eu,MartinezTorres:2011pr} (see Appendix~\ref{supp:energy_levels}).

Furthermore, the relative strength of the $Z_c(3900)$ couplings to the $D \bar{D}^{\ast}$ and $J/\psi\pi$ channels is an important quantity.
The BESIII Collaboration reported $\Gamma(Z_c(3900) \to D\bar{D}^\ast)/\Gamma(Z_c(3900) \to J/\psi\pi)= 6.2\pm 1.1\pm 2.7$ in Ref.~\cite{BESIII:2013qmu} using a Breit-Wigner parameterization, which is not appropriate because the $Z_c$ lies very close to the $D\bar{D}^\ast$ threshold~\cite{Chen:2015jgl}. Therefore, instead of using the decay-width ratio quoted in Ref.~\cite{BESIII:2013qmu}, we consider the event-yield ratio in the $Z_c$ region.
Specifically, for the data at $E=4.26$~GeV, we combine the efficiency-corrected event numbers for the $D\bar{D}^\ast$~\cite{BESIII:2015pqw} and $J/\psi\pi^+$ spectra~\cite{BESIII:2017bua} around the $Z_c$ mass in the window $(3900\pm 35)$~MeV, obtaining 
\begin{eqnarray}\label{eq.ratio_of_Zc_couplings_EX}
 R_{Z_c,\text{exp}} = \frac{\sum \text{events for} ~ Z_c \to D\bar{D}^\ast }{\sum \text{events for} ~ Z_c \to J/\psi\pi } = 8.07\pm 2.32\,,
\end{eqnarray}
which is taken into account as a constraint.

\section{Results and discussions} \label{pheno}

We perform a simultaneous fit to the data from both BESIII measurements~\cite{BESIII:2025qkn,BESIII:2022joj,BESIII:2015pqw,BESIII:2020qkh,BESIII:2023wqy,BESIII:2017bua} and lattice QCD calculations~\cite{Cheung:2017tnt,CLQCD:2019npr,Sadl:2024dbd}. The results of the best fit, denoted as Fit I, are presented in Appendix~\ref{supp:Fit_results}, with $\chi^2/\text{d.o.f.}=1.43$. For the event-yield ratio of $D\bar{D}^\ast$ and $J/\psi \pi$ near the $Z_c$, the theoretical result is $R_{Z_c,\text{th}} = 8.11 \pm 0.20$, which is consistent with the experimental value $R_{Z_c,\text{exp}} = 8.07\pm 2.32$ within uncertainties.

 
Poles are found on different Riemann sheets of the $J/\psi\pi(J/\psi \bar{K})$-$\bar{D}D^*(\bar{D}D^*_s)$ $T^{(s)}$-matrix defined in Eq.~\eqref{Tmatrix}, which are reached by analytically continuing the two-point loop function $G$ in Eq.~\eqref{eq.Gfunction}.
The ($\eta_1\eta_2$) Riemann sheet (RS) of the $T^{(s)}$-matrix is defined through the replacement
\begin{align}
G_i(t) & \rightarrow G_i(t)+\eta_i 2 i\rho_i(t)\,\label{eq:RiemannSheetDefinition}
\end{align}
where $\rho_i(t)=\lambda^{1/2}(t,m_{i1}^2,m_{i2}^2)/(16\pi t)$ denotes the two-body phase space in channel $i~(i=1,2)$. Here $t$ is the squared total c.m. energy of $J/\psi\pi(J/\psi\bar{K})$.
In this convention, the physical sheet is labeled as RS-I=(00), and the other three RSs are denoted as RS-II=(10), RS-III=(11), and RS-IV=(01).

\begin{figure}[tbh]
  \centering
  \includegraphics[width=\linewidth]{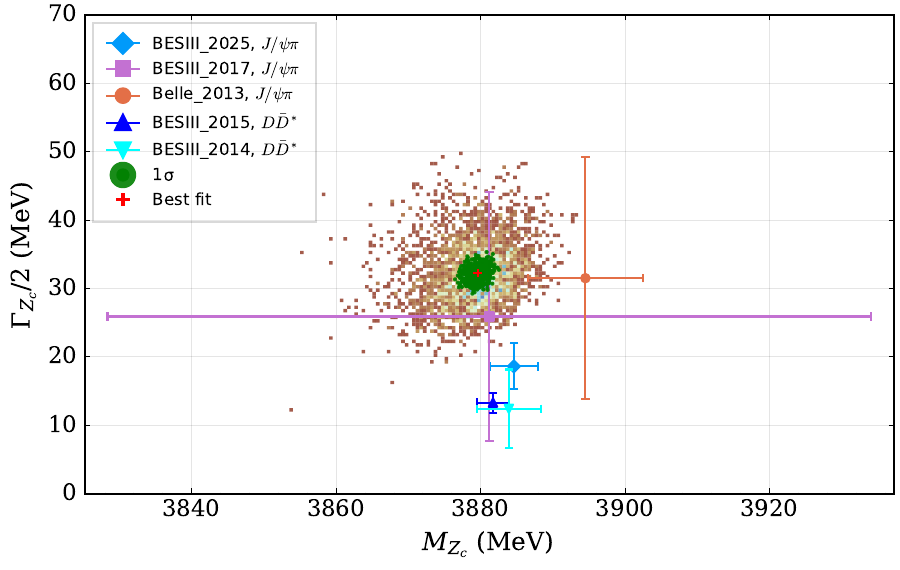}  
  \includegraphics[width=\linewidth]{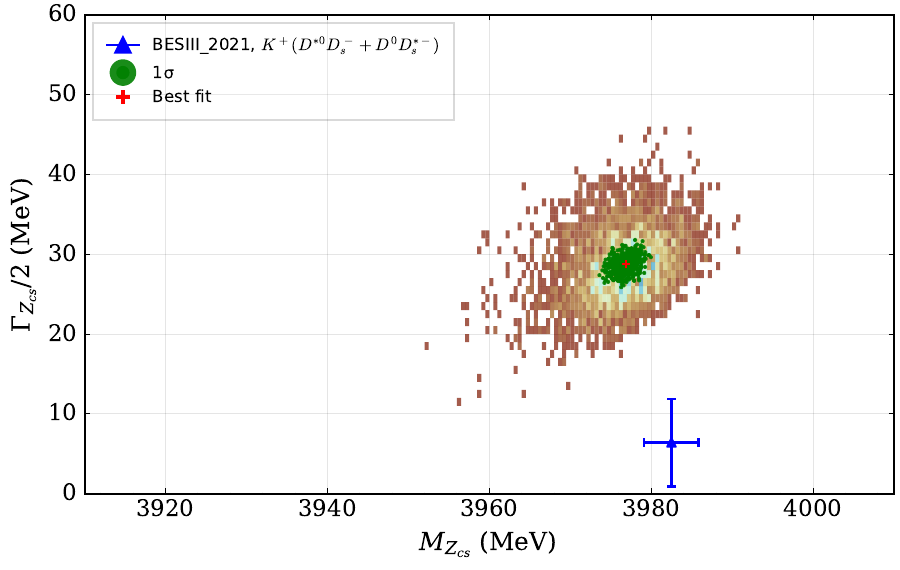}    
      \caption{$Z_c$ (top) and $Z_{cs}$ (bottom) pole positions in Fit I. The red crosses indicate the best-fit results. The green dots correspond to 640 parameter sets within 1$\sigma$, selected from $2\times 10^9$ randomly sampled parameter sets following a multivariate normal distribution that includes the parameter correlations (the brown dots denote poles from $3\times10^3$ such parameter sets). The $Z_c(3900)$ mass and half-width from the BESIII~\cite{BESIII:2025qkn,BESIII:2017bua,BESIII:2013qmu,BESIII:2015pqw} and Belle~\cite{Liu:2013dau} analyses, and the $Z_{cs}(3985)$ mass and half-width from the BESIII~\cite{BESIII:2020qkh} analysis, are shown for comparison.}
    \label{fig.ZcZcspole}
\end{figure}

To propagate the statistical uncertainties from data to the physical results, we generate $2\times 10^9$ parameter sets following a multivariate normal distribution that incorporates the parameter correlations. Among them, 640 sets lie within 1$\sigma$. For all of these 1$\sigma$ parameter sets, there is a pole on RS-III in the $J/\psi\pi(J/\psi \bar{K})$-$\bar{D}D^*(\bar{D}D^*_s)$ coupled-channel $T^{(s)}$-matrix above the $D\bar{D}^{\ast}_{(s)}$ threshold, which we identify as the resonance corresponding to the $Z_c(3900)^\pm(Z_{cs}(3985)^\pm)$. The pole distributions are shown by the green dots in Fig.~\ref{fig.ZcZcspole}.
The parameters of the $J/\psi\pi(J/\psi \bar{K})$-$\bar{D}D^*(\bar{D}D^*_s)$ $T^{(s)}$-matrix, together with the pole masses and half-widths of the $Z_c(3900)$ and $Z_{cs}(3985)$, are given in Table~\ref{tab:poles}.

\begin{table*}[tbh]
\caption{Parameters of the $J/\psi\pi(J/\psi \bar{K})$-$\bar{D}D^*(\bar{D}D^*_s)$ $T^{(s)}$-matrix in Eq.~\eqref{Tmatrix}, together with the pole masses and half-widths of the $Z_c(3900)$ and $Z_{cs}(3985)$ determined in Fit I.}
\begin{tabular}{ l  c  c  c c c    cc}
\hline\hline
\multirow{2}{*}{}  & \multirow{2}{*}{$C_{12}$~[fm$^2$] } &  \multirow{2}{*}{$C_{1Z}$~[fm$^2$]} &  \multirow{2}{*}{$b$~[fm$^3$] }& \multicolumn{2}{c}{$Z_c$ [MeV]} &  \multicolumn{2}{c}{$Z_{cs}$ [MeV]}   \\
\cline{5-8}
&&&& {\rm Mass} & $\Gamma/2$ & {\rm Mass} & $\Gamma/2$ \\
\hline
 & $-0.072\pm 0.003$ &  $-0.200 \pm 0.004$ &  $-0.229 \pm 0.024$ & $3879.6 \pm 4.8$ & $32.2 \pm 4.7$ & $3976.9 \pm 5.1$ & $28.8 \pm 5.9$  \\
\hline
\hline
\end{tabular}
\label{tab:poles}
\end{table*}

\begin{figure}[tbh]
  \centering
  \includegraphics[height=2.5cm,width=8.5cm]{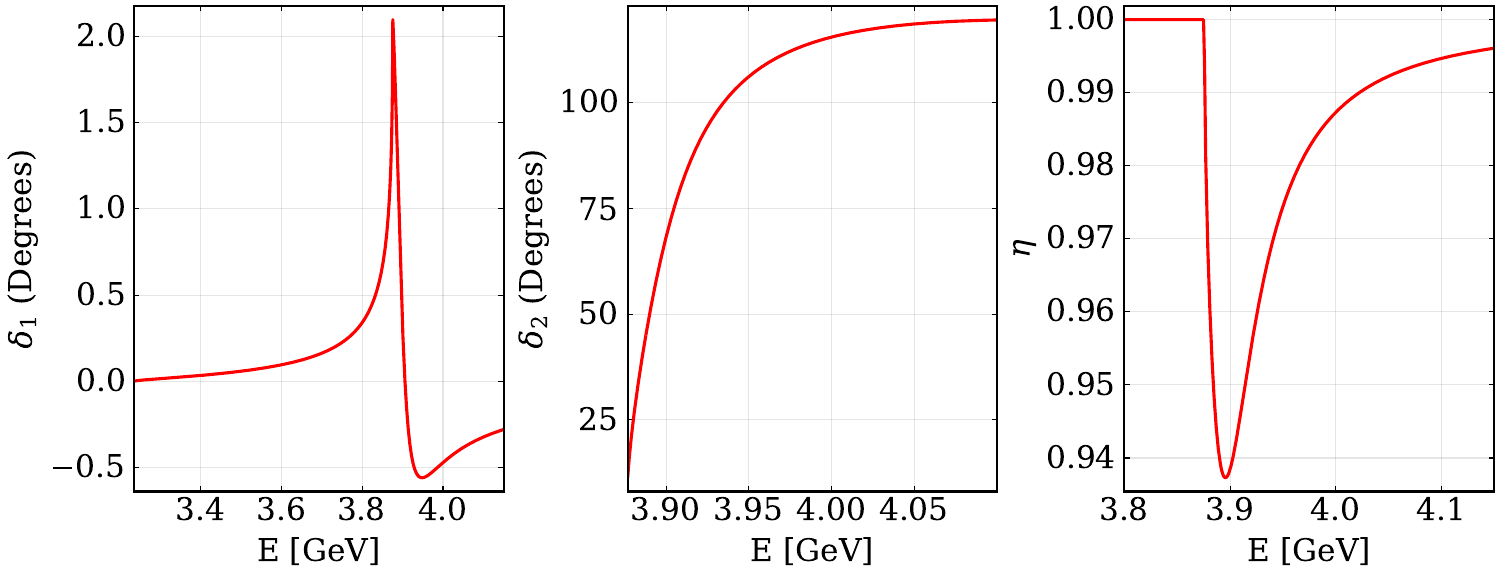}
  \includegraphics[height=2.5cm,width=8.5cm]{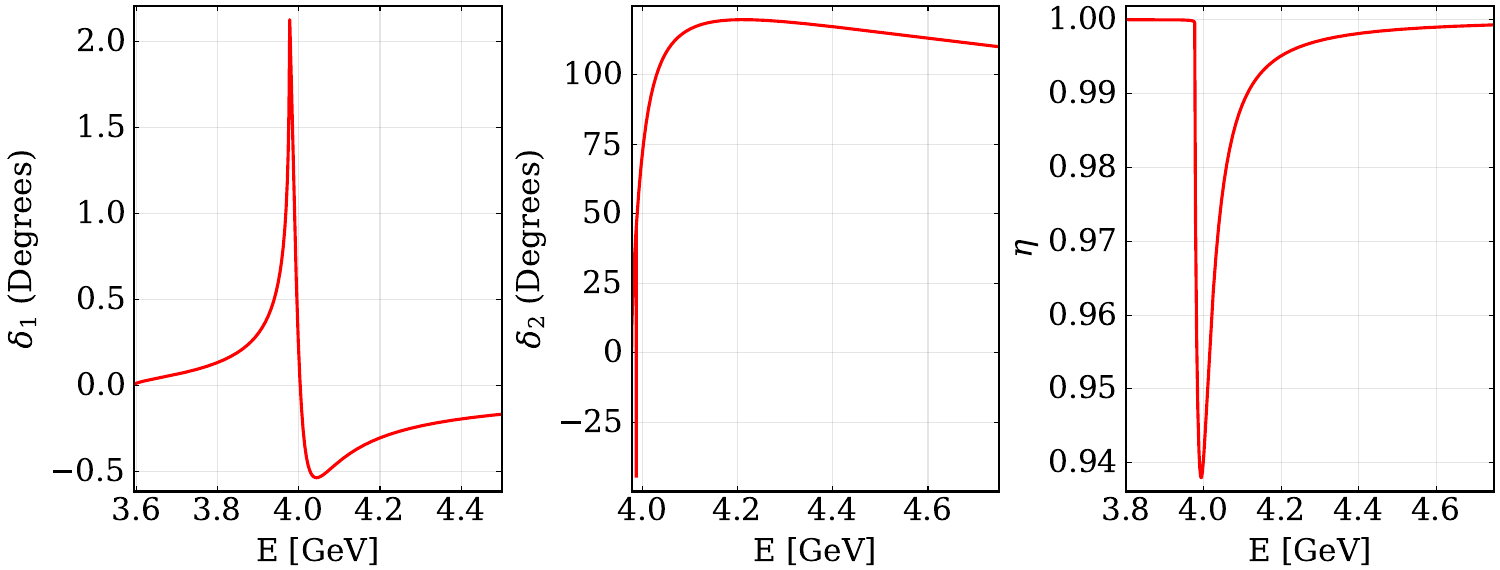}
      \caption{Theoretical predictions in Fit I for the phase shifts of $J/\psi\pi \to J/\psi\pi$ (top left), $\bar{D}D^\ast \to \bar{D}D^\ast$ (top center), and the corresponding inelasticity (top right). The bottom panels show the corresponding predictions for $J/\psi K \to J/\psi K$ (left), $\bar{D}D_s^\ast \to \bar{D}D_s^\ast$ (center), and the inelasticity (right).}
    \label{fig.phaseshiftinelasticity}
\end{figure}

The effects of poles on unphysical Riemann sheets are reflected in the phase shifts $\delta_j$ and inelasticities $\eta$ of the coupled-channel scattering amplitudes, 
which are defined by the $S$-wave $S$-matrix elements as $S_{jj}=\eta e^{2i\delta_j}$. 
In Fig.~\ref{fig.phaseshiftinelasticity}, we present our predictions for the phase shifts and inelasticities in
the $J/\psi\pi(J/\psi \bar{K})$-$\bar{D}D^*(\bar{D}D^*_s)$ coupled-channel
scattering. One observes that both $\delta_2$ in $J/\psi\pi$-$\bar{D}D^*$ and in $J/\psi \bar{K}$-$\bar{D}D^*_s$ coupled-channel
scattering pass through $90^\circ$, which is a signature of the pole on RS-III.

There have been suggestions that the BESIII data on $Z_c(3900)$ may be dominated by triangle singularities~\cite{Pilloni:2016obd,Yu:2024sqv}, which would diminish the necessity of introducing a $Z_c$ pole to describe the data. 
Thus, to clarify the situation, it is essential to perform additional fits. In Fit II we neglect the triangle diagrams, while in Fit III we remove the pole contribution by replacing the $T^{(s)}$-matrix with a constant matrix. 
The results of Fits II and III, with $\chi^2/\text{d.o.f.}=1.57$ and $2.50$, respectively, are presented in Appendix~\ref{supp:Fit_results}.
Fit II can reproduce the main features of the data, except for the $K^+K^-$ and $J/\psi K^\pm$ invariant mass spectra in
$e^+e^- \to J/\psi K^+ K^-$ at $E=4.68$~GeV and the finite-volume energy levels reported by M.~Sadl et al.~\cite{Sadl:2024dbd}. 
By contrast, in Fit III the peak around 3.9 GeV in the $J/\psi\pi^\pm$ mass spectra and the peak around 3.98 GeV in the
$K^+$ recoil-mass distribution and the $J/\psi K^\pm$ mass spectra at $E=4.68$~GeV cannot be reproduced. Therefore, we conclude that the pole contribution is indispensable for describing the peaks in the mass spectra, while the triangle diagrams also improve the overall fit quality.

The best fit constrains the ratios of the $Z_c(Z_{cs})$ couplings to the $D\bar D^*(D_s\bar{D}^\ast+D\bar{D}_s^\ast)$ and $J/\psi \pi(J/\psi K)$ channels.
From the residues of the $T$-matrix and $T^{(s)}$-matrix elements, we obtain in Fit I the ratio of the $Z_c$ couplings to the $D\bar D^*$ and $J/\psi \pi$ channels, $|g_{Z_cD\bar D^*}/g_{Z_cJ/\psi\pi}| =8.58\pm 0.20$, and the corresponding ratio for the $Z_{cs}$ couplings to the $(D_s\bar{D}^\ast+D\bar{D}_s^\ast)$ and $J/\psi K$ channels, $|g_{Z_{cs}  (D_s\bar{D}^\ast+D\bar{D}_s^\ast)}/g_{Z_{cs} J/\psi K}| =8.59\pm 0.19$. The remarkable consistency between these ratios strongly supports the interpretation of the $Z_c(3900)$ and $Z_{cs}(3985)$ as SU(3) flavor partners.

\begin{table*}[tbh]
\caption{The isoscalar $D\bar D^* (D_s\bar{D}^\ast+D\bar{D}_s^\ast)$ $S$-wave scattering length $a_0$, effective range $r_0$, and compositeness coefficient $\bar X_A$ determined in Fit I.}
\begin{tabular}{ l  c  c  c }
\hline\hline
& {$a_0$~[fm] } & {$r_0$~[fm]} &  {$\bar X_A$ }   \\
\hline
$Z_c(3900)$  &  $0.89\pm 0.13 +i(0.04\pm 0.01)$ &  $-1.34\pm 0.17 +i(0.02\pm 0.01)$  & $0.50\pm 0.04$ \\
\hline
$Z_{cs}(3985)$ &  $1.03\pm 0.17 +i(0.05\pm 0.02)$ &  $-1.27\pm 0.17 +i(0.02\pm 0.01)$  & $0.54\pm 0.05$ \\
\hline
\hline
\end{tabular}
\label{tab:compositeness}
\end{table*}

To investigate the nature of the $Z_c(3900)$ and $Z_{cs}(3985)$, we calculate their compositeness coefficients following the approach in Ref.~\cite{Matuschek:2020gqe}.
Using the isoscalar $D\bar D^* (D_s\bar{D}^\ast+D\bar{D}_s^\ast)$ $S$-wave scattering length $a_0$ and effective range $r_0$, the compositeness coefficient $\bar X_A$, which measures the probability of finding the two-body component in a near-threshold resonance, is expressed as~\cite{Matuschek:2020gqe}
\begin{equation}
\bar X_A =(1+2|r_0/a_0|)^{-1/2} \,.
\end{equation}
The low-energy parameters $a_0$ and $r_0$ are extracted from the effective-range expansion of the scattering amplitude
\begin{equation}\label{eq.EREofTmatrix}
\frac1{T_{22}^{(s)}(k)} = - \frac1{8\pi \sqrt{t}} \left[\frac1{a_0} + \frac12 r_0 k^2 -i\, k + \mathcal{O}(k^4) \right],
\end{equation}
with $k$ the c.m. momentum. The results in Table~\ref{tab:compositeness} again show consistency between the two states and further support their interpretation as flavor partners. The values of
$\bar X_A^{Z_c(3900)}$ and $\bar X_A^{Z_{cs}(3985)}$, namely the probabilities of the $D\bar D^* (D_s\bar{D}^\ast+D\bar{D}_s^\ast)$ component in the
$Z_c(3900)$ and $Z_{cs}(3985)$ states, are both about 0.5, indicating that additional components are important in these states.
Clarifying the nature of these additional components requires more specific theoretical models.

\section{Conclusions} \label{conclu}

This work presents a comprehensive analysis of the $Z_c(3900)$ and $Z_{cs}(3985)$ states through a simultaneous description of experimental and lattice QCD data. The BESIII data included in the fit comprise the $\pi^+\pi^-$ and
$J/\psi\pi^\pm$ invariant mass distributions for $e^+e^-
\rightarrow J/\psi \pi^+\pi^-$~\cite{BESIII:2025qkn}, the $K^+ K^-$ invariant mass spectra for
$e^+e^- \to J/\psi K^+ K^-$ at 17 $e^+e^-$ c.m. energies from $E=4.1271$ to
$4.3583$~GeV~\cite{BESIII:2022joj}, the $D^0 D^{\ast-}$ mass spectra for $e^+e^-
\rightarrow D^0 D^{\ast-} \pi^+$ at $E=4.23$~GeV and $4.26$~GeV~\cite{BESIII:2015pqw},
the $K^+$ recoil-mass distribution for $e^+e^- \rightarrow (D^{\ast 0} D_s^{-}+D^0 D_s^{\ast -}) K^+$~\cite{BESIII:2020qkh}, the $K^+K^-$ and $J/\psi K^\pm$ invariant mass spectra for $e^+e^- \to J/\psi K^+ K^-$ at $E=4.68$~GeV~\cite{BESIII:2023wqy},
and the efficiency-corrected event-yield ratio of the $D\bar{D}^\ast$ and $J/\psi\pi^+$ channels around the $Z_c$ mass~\cite{BESIII:2015pqw,BESIII:2017bua}.
The framework incorporates three essential ingredients: open-charm triangle diagrams with TS contributions, $J/\psi\pi(J/\psi \bar{K})$-$\bar{D}D^*(\bar{D}D^*_s)$ coupled-channel interactions, and the strong $\pi\pi$-$K\bar K$ FSI in $e^+e^-\to J/\psi \pi^+\pi^-$ and $J/\psi K^+ K^-$. The analysis also includes finite-volume energy levels from lattice QCD simulations~\cite{Cheung:2017tnt,CLQCD:2019npr,Sadl:2024dbd}, which provide complementary constraints on the $Z_c(3900)$ properties. 

Our best fit successfully describes the experimental and lattice data. In this fit, the $Z_c(3900)$ and $Z_{cs}(3985)$ poles lie above the $D\bar{D}^*$ and $D_s\bar{D}^*+D\bar{D}_s^*$ thresholds, respectively, indicating that both states are resonances. This analysis leads to a robust and precise extraction of the corresponding pole positions, with the $Z_c(3900)$ at $(3879.6 \pm 4.8)$~MeV and half-width $(32.2 \pm 4.7)$~MeV, and the $Z_{cs}(3985)$ at $(3976.9 \pm 5.1)$~MeV and half-width $(28.8 \pm 5.9)$~MeV.
We also perform two auxiliary fits by omitting either the triangle diagrams or the $Z_c(Z_{cs})$ pole contributions. We find that the $Z_c(Z_{cs})$ pole contribution is indispensable for describing the data, while the triangle diagrams also play a non-negligible role. The best fit supports the interpretation of the $Z_c(3900)$ and $Z_{cs}(3985)$ as SU(3) flavor partners within the same SU(3) light flavor octet.
The extracted ratio of the $Z_c$ couplings to the $D\bar D^*$ and $J/\psi \pi$ channels and the corresponding ratio of the $Z_{cs}$ couplings to the $D_s\bar{D}^\ast+D\bar{D}_s^\ast$ and $J/\psi K$ channels are remarkably consistent: $|g_{Z_cD\bar D^*}/g_{Z_cJ/\psi\pi}| =8.58\pm 0.20$ and $|g_{Z_{cs}  (D_s\bar{D}^\ast+D\bar{D}_s^\ast)}/g_{Z_{cs} J/\psi K}| =8.59\pm 0.19$.
The compositeness coefficients of the $Z_c(3900)$ and $Z_{cs}(3985)$ indicate sizable $D\bar{D}^*$ and $D_s\bar{D}^*+D\bar{D}_s^*$ components, respectively, while additional short-distance contributions also play an important role in their structure. Further investigation of these additional components based on the results of our analysis would provide valuable insights into the nature of these exotic hadrons.

\begin{acknowledgements}

We are grateful to Zhi-Hui Guo and Bing Wu for helpful discussions. We acknowledge Zhen-Tian Sun for providing
us with the data in Ref.~\cite{BESIII:2025qkn}. This work is supported in part by the National Natural Science Foundation of China (NSFC) under Grants No.~12361141819, No. 12125507, No. 12447101, and No. 12547111; by the Fundamental Research Funds for the
Central Universities; by the National Key R\&D Program of China under Grant No. 2023YFA1606703; and
by the Chinese Academy of Sciences under Grant No. YSBR-101.

\end{acknowledgements}

\medskip
\medskip
\medskip

\begin{appendix}

\section{$J/\psi\pi(J/\psi \bar{K})$-$\bar{D}D^*(\bar{D}D^*_s)$ coupled-channel interaction
$T^{(s)}$-matrix}
\label{supp:Ts_matrix}

\begin{figure}[tb]
\centering
\includegraphics[width=10cm]{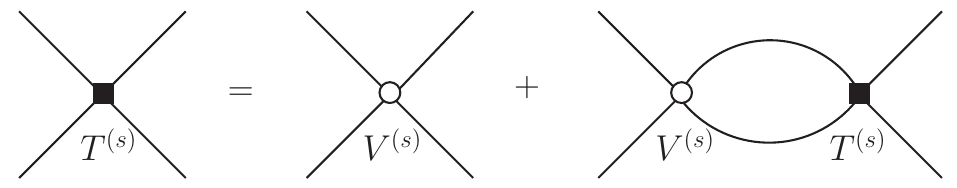}
  \caption{ Diagrammatic representation of the $J/\psi\pi(J/\psi \bar{K})$-$\bar{D}D^*(\bar{D}D^*_s)$ unitary scattering amplitude $T^{(s)}$ given in Eq.~\eqref{Tmatrix}, which can be rewritten as $T^{(s)} = V^{(s)} + V^{(s)} G^{(s)}T^{(s)}$.
    }
  \label{fig.FeynmanDiagramTmatrix}
\end{figure}

By denoting the $J/\psi\pi (J/\psi \bar{K})$ and $ D\bar{D}^\ast (\bar{D}D^*_s)$ with channel 1 and 2,
respectively, the $J/\psi\pi(J/\psi \bar{K})$-$\bar{D}D^*(\bar{D}D^*_s)$ coupled-channel rescattering
$T^{(s)}$-matrix can be obtained as
\begin{equation}\label{Tmatrix}
T^{(s)} = (\mathbb{I}-V^{(s)} \cdot G^{(s)})^{-1} \cdot V^{(s)}~,
\end{equation}
where the superscript $s$ denotes the strange sector and the diagrammatic representation is shown in Fig.~\ref{fig.FeynmanDiagramTmatrix}.
The $J/\psi \pi(J/\psi K) \to J/\psi \pi(J/\psi K)$ interaction is known to be weak~\cite{Yokokawa:2006td,Liu:2012dv, Yan:2026oil} and can be neglected by setting $V_{11}^{(s)}=0$.
The other matrix elements of the potential can be written as
\bea\label{eq:Vij}
V_{12}^{(s)} &=& 4\sqrt{m_{D_{(s)}}m_{D^*}m_{\psi}m_{\pi}} \tilde{V}_{12}^{(s)}\,,\nn\\
 V_{22}^{(s)} &=& 4m_{D_{(s)}}m_{D^*}\tilde{V}_{22}^{(s)}\,.
\eea
Heavey quark spin symmetry (HQSS) and SU(3) flavor symmetry imply  $\tilde{V}^s_{12}=\tilde{V}_{12}$ and $\tilde{V}_{22}^s=\tilde{V}_{22}$.
Note that the nonrelativistic normalization factors $\sqrt{m_D}$ or $\sqrt{m_{D_s}}$ in Eq.~\eqref{eq:Vij} introduces a small SU(3) breaking effection.
For the $ D\bar{D}^\ast \to J/\psi \pi (D\bar{D}_s^\ast \to J/\psi K)$ $S$-wave interaction, we use
the simplest assumption of a constant coupling:
\bea
\label{eq:V12}
\tilde{V}^{(s)}_{12}= C_{12}\,.
\eea
For the $ D\bar{D}^\ast \to  D\bar{D}^\ast (D\bar{D}_s^\ast \to  D\bar{D}_s^\ast)$ scattering, we allow an energy-dependent term~\cite{Albaladejo:2015lob,Du:2022jjv}
\bea
\label{eq:V22}
\tilde{V}^{(s)}_{22} = C_Z + \frac{b}{2(m_{D_{(s)}}+m_{D^*})}\left[ t-(m_{D_{(s)}}+m_{D^*})^2\right]\,,
\eea
where $t$ is the square of the total c.m. energy of $D\bar D^* (D_s\bar D^*)$, and $C_{1Z}$ and $b$ are two parameters.
In Eq.~\eqref{Tmatrix}, $G^{(s)}$ is the two-point loop function diagonal matrix 
$G=\text{diag}\{G_1(t),G_2(t)\}$. 
We use the dimensionally regularized two-point scalar loop function~\cite{Oller:1998zr},
\begin{eqnarray}\label{eq.Gfunction}
G_i(t)&=&i\int\frac{d^4q}{(2\pi)^4}\frac{1}{(q^2-m_{i1}^2+i\epsilon)\left[(p-q)^2-m_{i2}^2+i\epsilon\right]}\nn\\
&=&\frac1{16\pi^2}\bigg\{a_i(\mu)+\log\frac{m_{i1}^2}{\mu^2}+\frac{m_{i2}^2-m_{i1}^2+t}{2t} \log\frac{m_{i2}^2}{m_{i1}^2} \nonumber \\ 
&&+\frac{q_i}{\sqrt{t}}\log\frac{(t+2q_i \sqrt{t})^2-(m_{i1}^2-m_{i2}^2)^2}{(t-2q_i \sqrt{t})^2-(m_{i1}^2-m_{i2}^2)^2}\bigg\},
\end{eqnarray}
where $m_{i,n}$ denotes the mass of the $n$th particle in
channel $i$, $t=p^2$, and $q_i^2 = \lambda(t,m_{i1}^2,m_{i2}^2)/(4t)$ represents the c.m. momentum squared for channel $i$. Variations in the renormalization scale $\mu$ can be compensated by a corresponding change of the subtraction constants $a_i(\mu)$. The values of $a_1(\mu=1\, \text{GeV}) = -2.77$ and $a_2(\mu=1\, \text{GeV})=-3.0$ are fixed following~\cite{Du:2022jjv}.

\section{ Description of the finite volume energy levels }
\label{supp:energy_levels}

To describe the lattice energy levels of $Z_c(3900)$, we employ the method proposed in Refs.~\cite{Doring:2011vk,Doring:2012eu}. For a cubic box of side $L$ with periodic boundary conditions, the finite-volume correction of the $G$ function in the c.m. frame is given by~\cite{Doring:2011vk,Doring:2012eu,MartinezTorres:2011pr,Yan:2023bwt}
\begin{align}\label{eq.deltaG}
\Delta G_i(t)= \frac{1}{L^3} \sum_{\vec{n}}^{|\vec{q}|<q_{\rm max}} I(|\vec{q}|)-\int^{|\vec{q}|<q_{\rm max}} \frac{d^3\vec{q}}{(2\pi)^3} I(|\vec{q}|)  \,,
\end{align}
where
\begin{eqnarray}\label{eq.Ifunctionetc}
\vec{q}&=& \frac{2\pi}{L} \vec{n},\,(\vec{n} \in \mathbb{Z}^3) \,, \quad \omega_{ij} =\sqrt{|\vec{q}|^2+m_{ij}^2}\,, \nn\\
I(|\vec{q}|)&=&\frac{\omega_{i1}+\omega_{i2}}{2\omega_{i1} \omega_{i2} \,[t-(\omega_{i1}+\omega_{i2})^2]} \,.
\end{eqnarray}
In the practical calculation, we adopt $q_{\rm max}=\frac{2\pi}{L}n_{\rm max}$ with $n_{\rm max}=20$. 
The finite-volume loop function $G$ in Eq.~\eqref{Tmatrix} is be replaced by
\begin{eqnarray}\label{eq.tildeG}
\tilde{G}_i(t) = G_i(t)+\Delta G_i(t)  \,,
\end{eqnarray}
where $G_i(t)$ and $\Delta G_i(t)$ are given in Eqs.~\eqref{eq.Gfunction} and~\eqref{eq.deltaG}, respectively.
In this method, the finite-volume effects are included via the $G$ function, while finite-volume corrections to the potential $V$ in Eq.~\eqref{Tmatrix} are exponentially suppressed and thus neglected.
The discrete energy levels in the box correspond to poles of the finite-volume $T$-matrix in Eq.~\eqref{Tmatrix}, obtained by solving 
\begin{eqnarray}\label{eq.detfv}
 \det{[\mathbb{I} - V \cdot \tilde{G}(s)]} =0\,.
\end{eqnarray}

\section{Fit results}
\label{supp:Fit_results}
\begin{figure*}[tbhp]
  \centering
     \includegraphics[height=20cm,width=\linewidth]{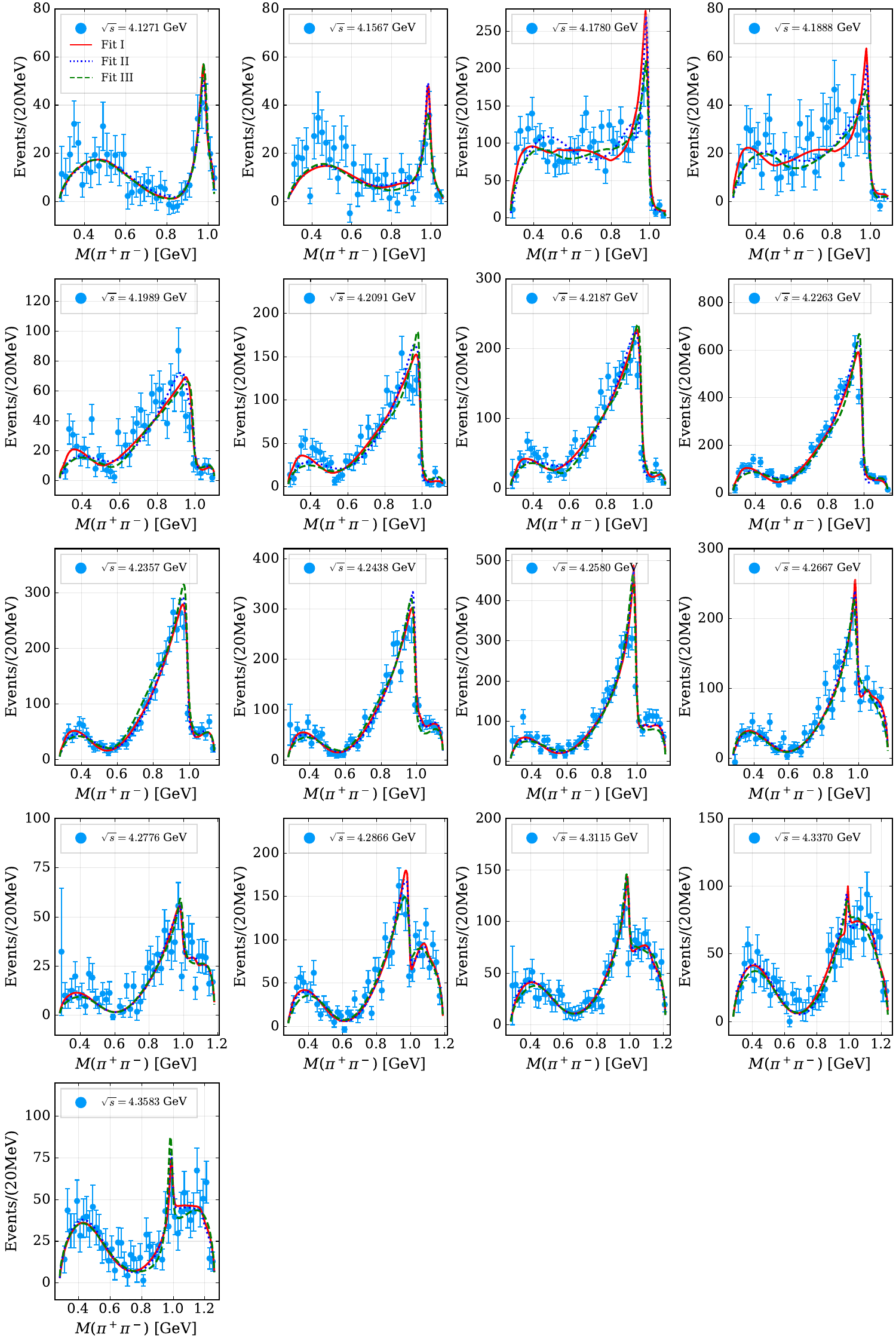}
\caption{ Fit results of the $\pi^+\pi^-$ invariant mass spectra in
 $e^+e^- \to J/\psi \pi^+\pi^-$ at 17 $e^+e^-$ c.m. energies from $E=4.1271$ to
$4.3583$~GeV~\cite{BESIII:2025qkn} for Fits I (solid), II (dotted), and III (dashed). 
}    
   \label{fig.FitResults_pipi}
\end{figure*}

\begin{figure*}[tbhp]
  \centering
     \includegraphics[height=20cm,width=\linewidth]{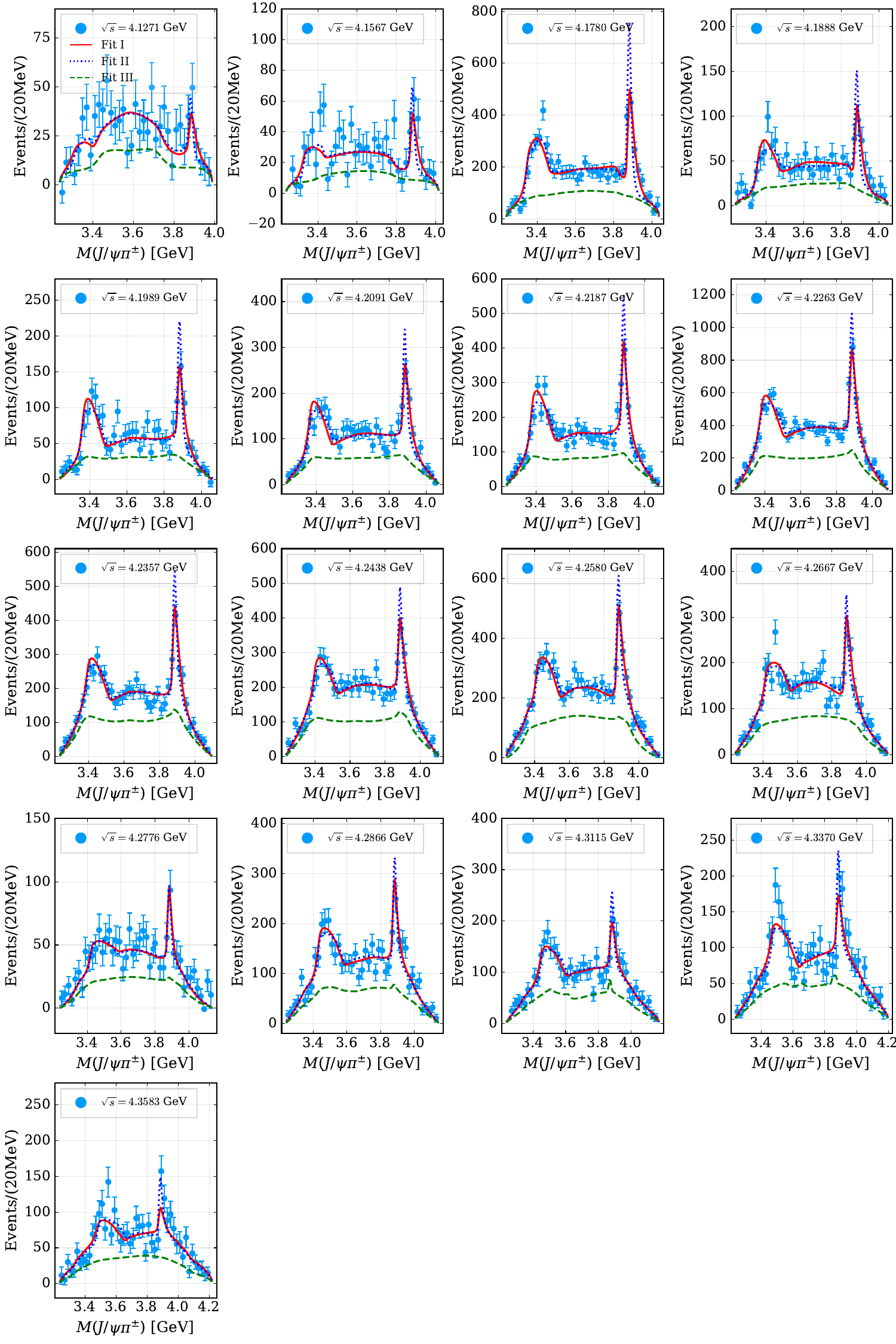}
\caption{ Fit results of the $J/\psi\pi^\pm$ invariant mass spectra in
 $e^+e^- \to J/\psi \pi^+\pi^-$ at 17 $e^+e^-$ c.m. energies from $E=4.1271$ to
$4.3583$~GeV~\cite{BESIII:2025qkn} for Fits I (solid), II (dotted), and III (dashed).  
}    
   \label{fig.FitResults_psipi}
\end{figure*}

\begin{figure*}[tbhp]
  \centering
     \includegraphics[height=20cm,width=\linewidth]{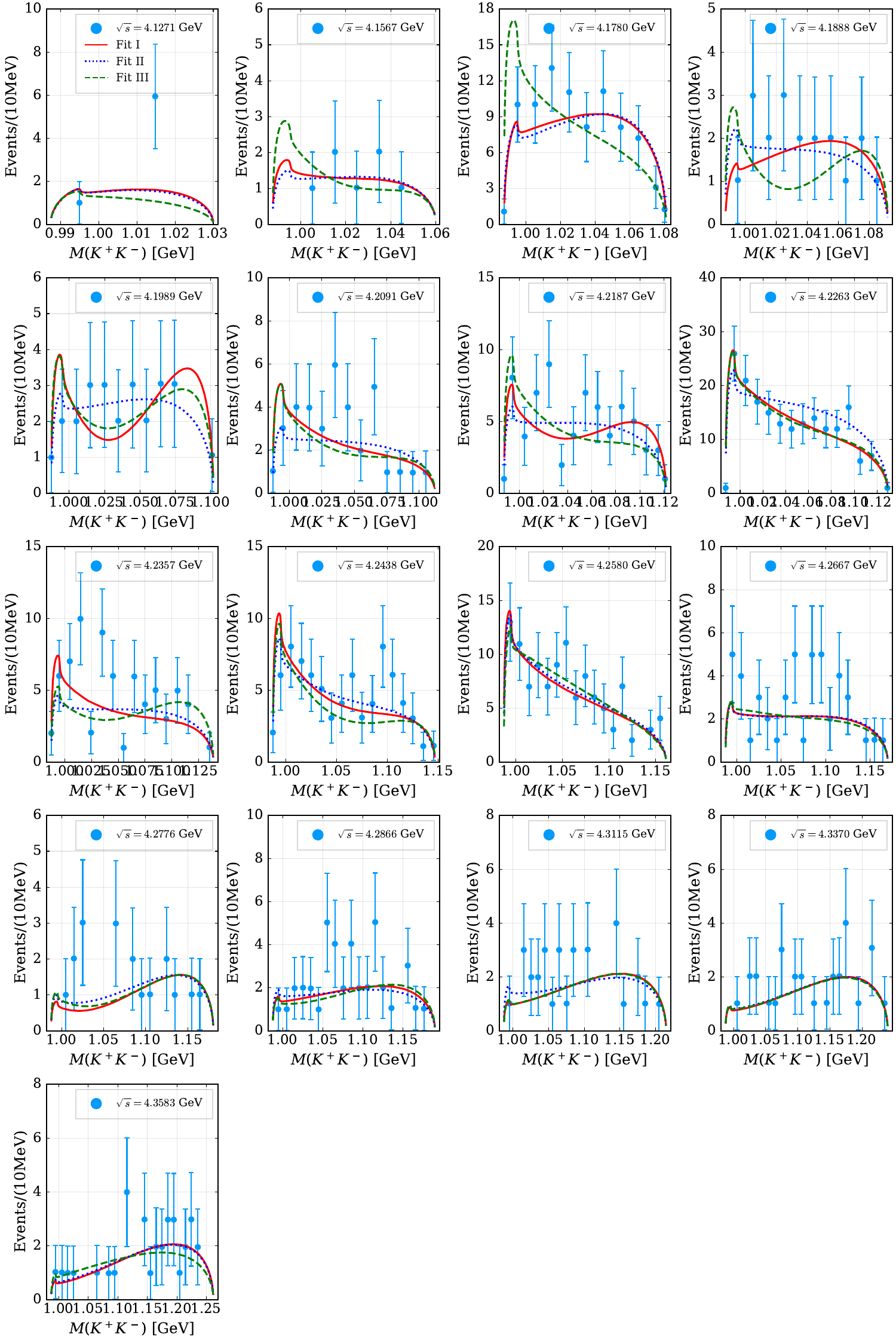}
\caption{ Fit results of the $K^+ K^-$ invariant mass spectra in
 $e^+e^- \to J/\psi K^+ K^-$ at 17 $e^+e^-$ c.m. energies from $E=4.1271$ to
$4.3583$~GeV~\cite{BESIII:2022joj} for Fits I (solid), II (dotted), and III (dashed). 
}    
   \label{fig.FitResults_KK}
\end{figure*}

\begin{figure*}[tbh]
  \centering
     \includegraphics[height=5cm,width=5cm]{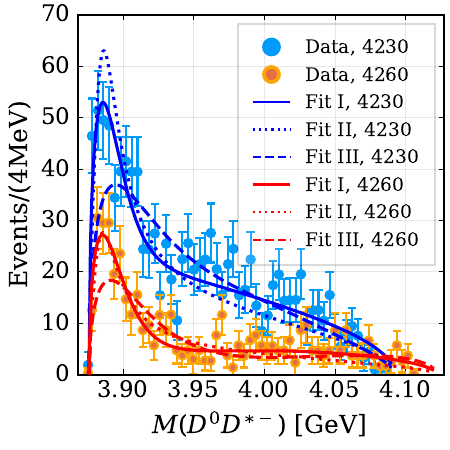}
     \includegraphics[width=\linewidth]{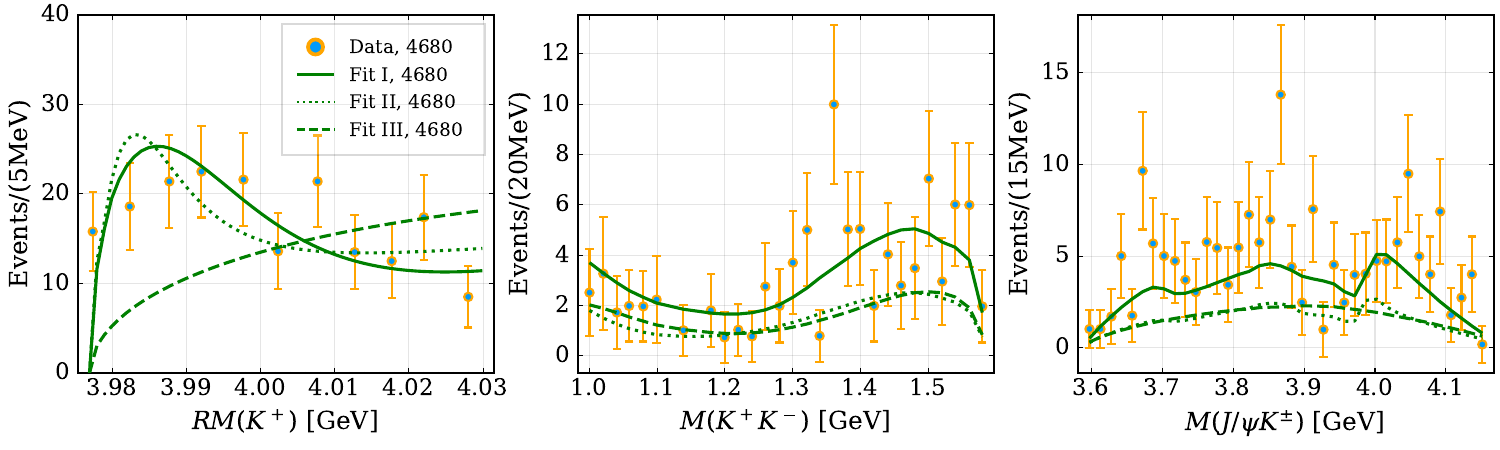}
     \includegraphics[width=\linewidth]{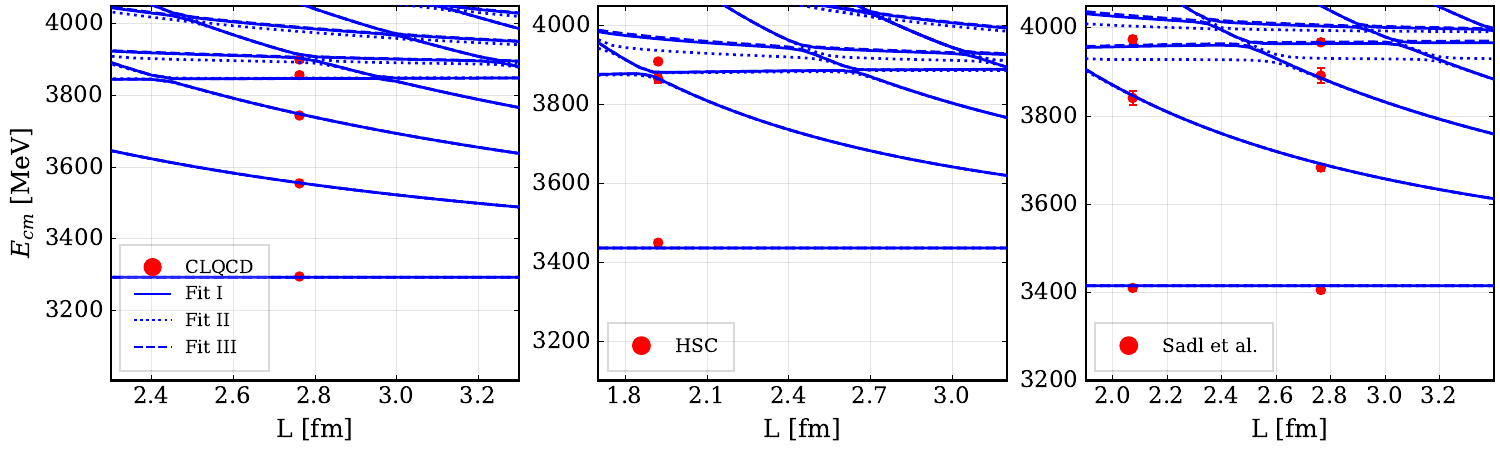}
\caption{Fit results and the experimental and lattice data. The results of Fits I, II, and III are shown as the solid, dotted, and dashed lines, respectively. Top panel: the fit results in comparison with the data of the $D^0 D^{\ast-}$ mass spectra for $e^+e^-
 \rightarrow D^0 D^{\ast-} \pi^+$~\cite{BESIII:2015pqw} at the $e^+e^-$ c.m. energies $E=4.23$~GeV (blue) and $4.26$~GeV (red);
 middle panel: the fit results in comparison with the data of the $K^+$ recoil-mass distribution for $e^+e^-
 \rightarrow (D^{\ast 0} D_s^{-}+D^0 D_s^{\ast -}) K^+$~\cite{BESIII:2020qkh}; the $K^+K^-$ and $J/\psi K^\pm$ invariant mass spectra for
 $e^+e^- \to J/\psi K^+ K^-$~\cite{BESIII:2023wqy} (from left to right) at the $e^+e^-$ c.m. energies $E=4.68$~GeV (green); bottom panel:
 the predictions of the finite volume energy levels together with lattice data taken from the CLQCD Collaboration~\cite{CLQCD:2019npr}, the Hadron Spectrum Collaboration (HSC)~\cite{Cheung:2017tnt}, and Ref.~\cite{Sadl:2024dbd} (Sadl et al.), respectively.
}
   \label{fig.FitResults_dd_4680_lattice}
\end{figure*}

In this appendix, we present the fit curves for Fits~I, II, and III in comparison with the experimental and lattice data.
Figure~\ref{fig.FitResults_pipi} shows the $\pi^+\pi^-$ invariant mass distributions in $e^+e^- \to J/\psi \pi^+\pi^-$, and Fig.~\ref{fig.FitResults_psipi} the corresponding $J/\psi\pi^\pm$ spectra, each at 17 $e^+e^-$ c.m. energies from 4.1271 to 4.3583~GeV~\cite{BESIII:2025qkn}.
Figure~\ref{fig.FitResults_KK} gives the $K^+K^-$ spectra in $e^+e^- \to J/\psi K^+K^-$ over the same energy range~\cite{BESIII:2022joj}.
Figure~\ref{fig.FitResults_dd_4680_lattice} collects further comparisons: $D^0 D^{\ast-}$ mass spectra from $e^+e^- \to D^0 D^{\ast-}\pi^+$~\cite{BESIII:2015pqw}; strange-sector observables including the $K^+$ recoil-mass distribution for $e^+e^- \to (D^{\ast 0} D_s^{-}+D^0 D_s^{\ast -}) K^+$~\cite{BESIII:2020qkh} and the $K^+K^-$ and $J/\psi K^\pm$ invariant mass spectra for $e^+e^- \to J/\psi K^+K^-$ at $E=4.68$~GeV~\cite{BESIII:2023wqy}; and finite-volume energy-level predictions compared with lattice results from the CLQCD Collaboration~\cite{CLQCD:2019npr}, the Hadron Spectrum Collaboration~\cite{Cheung:2017tnt}, and Ref.~\cite{Sadl:2024dbd}.

\end{appendix}

\bibliography{refs}

\end{document}